\documentclass[12pt]{article}
\usepackage{hyperref}
\usepackage{amsmath}
\usepackage{graphicx}
\usepackage{amssymb} % Symbols for itemise

\def\be{\begin{equation}}
\def\ee{\end{equation}}
\def\bea{\begin{eqnarray}}
\def\eea{\end{eqnarray}}
\def\bi{\begin{itemize}}
\def\ei{\end{itemize}}
\def\dd{\mathrm{d}}

\date{}

\title{{\bf General reference frames and their associated space manifolds}}

\author{
Mayeul Arminjon\,$^{1,2}$ and Frank Reifler\,$^3$\\
$^1$ \small\it CNRS (Section of Theoretical Physics)\\
$^2$ \small\it Laboratory ``Soils, Solids, Structures, Risks'' (CNRS \& Universit\'es de Grenoble),\\
\small\it BP 53, F-38041 Grenoble cedex 9, France.\\
\small\it $^3$ Lockheed Martin Corporation, MS2 137-205,\\ 
\small\it 199 Borton Landing Road, Moorestown, New Jersey 08057, USA.
} % fin "author"

\begin{document}
\maketitle
\begin{abstract}
\noindent We propose a formal definition of a general reference frame in a general spacetime, as an equivalence class of charts. This formal definition corresponds with the notion of a reference frame as being a (fictitious) deformable body, but we assume, moreover, that the time coordinate is fixed. This is necessary for quantum mechanics, because the Hamiltonian operator depends on the choice of the time coordinate. Our definition allows us to associate rigorously with each reference frame F, a unique ``space" (a three-dimensional differentiable manifold), which is the set of the world lines bound to F. This also is very useful for quantum mechanics. We briefly discuss the application of these concepts to G\"odel's universe.

\end{abstract}

%%%%%%%%%%%%%%%%%%%%%%%%%%%%%%%%%%%%%%%%%%%%
\section{Introduction}\label{Intro}

While the notion of an inertial reference frame in a Minkowski spacetime  is easy to define, the notion of a general reference frame in a general spacetime is not. In the field of relativity and gravitation, it is of course often alluded to reference frames or to reference systems, in different contexts. In the literature on relativistic celestial mechanics and post-Newtonian calculations, this means essentially a {\it coordinate system,} thus mathematically a local chart $\chi$ defined in some open domain U of the spacetime manifold V. Then, a such coordinate system is usually ``attached'', in some sense, to a relevant astronomical body or system of bodies: e.g., to the Sun, to the barycenter of the solar system, to some planet, etc. (see e.g. Refs. \cite{Fock1964,Will93,DSX1991,RacineFlanagan2005}). However, if one changes the spatial coordinates in a way that does not depend on the time coordinate:
\be\label{internal-change}
x'^0=f((x^\mu)),\ x'^j=f^j((x^k)) \qquad (j,k=1,2,3),
\ee
then any element of matter which is ``at rest'' in the initial system, i.e., which has time-independent spatial coordinates in that system, will obviously remain ``at rest'' in the new system of coordinates. Thus, Landau \& Lifshitz \cite{L&L}, while studying the distances and time intervals in general relativity, considered that every ``physically admissible" spacetime coordinate system defines a reference frame, but that any coordinate change of the form (\ref{internal-change}) leaves us in the same reference frame. \\

The notion which remains implicit in most of this literature is that a general reference frame is some kind of fictitious fluid, thus is a fictitious body which is deformable in a general way, and that every fluid particle is represented by a world line in spacetime. (Note that spatial sections of spacetime are not truly relevant, because these are just sets of events, which do not last any time. Whereas, for a reference frame in the physical sense, there are ``point observers" attached with it, that exist at least for some open interval of time---thus world lines, really.) This notion, according to which a general reference frame corresponds with a ``three-dimensional congruence of world lines", was made explicit more than fifty years ago by Cattaneo \cite{Cattaneo}. As he noted, the physical admissibility of some coordinate system, which can be expressed by the condition \cite{Cattaneo}
\be\label{admissible-coordinates}
g_{00}>0, \qquad g_{jk}\dd x^j \dd x^k <0,
\ee
allows one to regard the coordinate lines 
\be\label{x^j = constant}
x^0 \mathrm{\ variable,\qquad }x^j =\mathrm{constant\ for\ }j=1,2,3
\ee
as the trajectories of the particles constituting the reference fluid. [We are using the $(+---)$ signature.] In fact, the condition (\ref{admissible-coordinates})$_1$ alone ensures that the tangent vector to any such line is time-like. Thus, within its own domain U, any admissible chart on the spacetime, $\chi: X \mapsto (x^\mu)$, defines a unique ``reference fluid'', given by its four-velocity field $v$: the components of $v$ in the chart $\chi$ are 
\be\label{Vmu}
v^0\equiv \frac{1}{\sqrt{g_{00}}}, \qquad v^j=0.
\ee
The vector (\ref{Vmu}) is invariant under the ``internal changes'' (\ref{internal-change}), provided that they satisfy $\partial x'^0/\partial x^0 >0$. Conversely, if we are given a unit (four-velocity) vector field $v$, then the coordinate systems {\it adapted} to the corresponding reference fluid are the ones in which $v$ has the form (\ref{Vmu}). Thus, one possible definition of a reference frame is: a time-like vector field \cite{RodriguesCapelas2007}.\\
 
We note that another definition of a ``frame'' is also used, notably in the metric-affine literature (see e.g. Refs. \cite{vdHeyde1975,Iliev1996,Hartley1995,MosnaPereira2004}): there, this name, or the more precise expression ``moving frame'', refers to a field of bases, thus it refers to a map $X \mapsto (u_\alpha(X))$, where $(u_\alpha(X))$ is a basis of the local tangent space $\mathrm{TV}_X$. Usually these bases are chosen to be orthonormal, in which case the field $(u_\alpha(X))$ is more commonly called a tetrad field. This other notion is also a useful one, of course. However, it is definitely different from the former notion of a reference frame as a three-dimensional congruence of world lines. In fact, if one of the normed basis vectors, say the first one $u_0$, is time-like, it alone suffices to define a reference frame F, characterized by its four-velocity field $v\equiv u_0$. Thus the other vectors of the basis, $u_j \ (j=1,2,3)$, are generally not ``attached'' to F, e.g. they may be rotating w.r.t. F.  \\

The present work originates from a study of the Hamiltonian operator H associated with the Dirac equation in a curved spacetime \cite{A42}. It was noted that this operator H is invariant only under purely spatial changes: 
\be\label{purely-spatial-change}
x'^0=x^0,\ x'^j=f^j((x^k)).
\ee 
This fact applies to any wave equation of relativistic quantum mechanics, provided \cite{A42} that its wave function $\psi$ transforms either as a scalar or as a four-vector.
\footnote{\ 
For the standard (Fock-Weyl) version of the Dirac equation in a curved spacetime, $\psi$ transforms as a quadruplet of scalar fields  (see e.g. Ref. \cite{BrillWheeler1957+Corr}). For two alternative versions of the Dirac equation in a curved spacetime (see Ref. \cite{A42} and references therein), $\psi$ transforms as a vector field.
}
In other words, the Hamiltonian operator depends on the reference frame {\it and} on the choice of the time coordinate. Therefore, in this paper, we formalize a more specific notion of reference frame, in which only purely spatial changes (\ref{purely-spatial-change}) are allowed: a reference frame will be essentially an {\it equivalence class of charts} modulo the relation (\ref{purely-spatial-change}), see the formal definition \hyperref[TheoremA]{below.} This formal definition allows us to associate rigorously with each reference frame, a unique ``space" M, which is a three-dimensional differentiable manifold. This turns out to be very useful---in our opinion, even necessary---in the discussion of the Hamiltonian operator, the Hilbert space scalar product, and the hermiticity of the Hamiltonian \cite{A42}. Indeed, the axioms of quantum mechanics need to introduce a space of ``states'', each of which is a function depending only on the space coordinates: $\psi=\psi((x^j))$. As is well known, it is on this space of states $\mathcal{H}$ that the operators of quantum mechanics, e.g. the Hamiltonian H, are acting. (Of course they are unbounded operators, i.e. they are defined on subspaces of $\mathcal{H}$.) In the absence of an intrinsic notion of a space manifold, the very space of states $\mathcal{H}$ would depend on the choice of the coordinate system, which is hardly acceptable. In contrast, once we have a space manifold M, the states can be defined intrinsically as functions of the point $x \in \mathrm{M}$. To our knowledge, these results are new. In particular, except for allusions in previous works by one of us \cite{A16,A35}, we are not aware of another work associating a unique three-dimensional space manifold with a general reference frame in a general spacetime.

\section{Definition of a reference frame F}\label{Def-F}

Equation (\ref{purely-spatial-change}) introduces a relation between two coordinate systems or charts $\chi$ and $\chi '$ on the spacetime V. That relation involves the transition map $F\equiv \chi '\circ \chi ^{-1}$, with $F(\mathbf{X})=\mathbf{X'}\equiv (x'^\mu )\equiv (F^\mu(\mathbf{X}))$ and $\mathbf{X}\equiv (x^\mu )$. Note that the domain (of definition) of the map $F$ is the very {\it intersection} of the domains of the two charts, more precisely it is $\chi(\mathrm{\mathrm{Dom}(\chi )\cap \mathrm{Dom}(\chi' )})$. Therefore, to define an equivalence relation between charts based on Eq. (\ref{purely-spatial-change}), we must limit ourselves to charts whose domains all contain some open subset $\mathrm{U}\subset \mathrm{V}$: \\

\paragraph{Theorem A.}\label{TheoremA} {\it Let $\mathrm{V}$ be a differentiable manifold of dimension $N+1$ ($N\geq 1$) and let $\mathcal{A}$  be the atlas that defines the manifold structure of $\mathrm{V}$. For any open set $\mathrm{U}\subset \mathrm{V}$, define the subset $\mathcal{A}_\mathrm{U}\equiv \{\chi \in \mathcal{A};\ \mathrm{Dom}(\chi )\supset \mathrm{U}\}$ of the atlas $\mathcal{A}$. Let the open set $\mathrm{U}$ be such that $\mathcal{A}_\mathrm{U}$ is non-empty and, for any two charts $\chi ,\chi ' \in \mathcal{A}_\mathrm{U}$, set 
\bea\label{R_U}
\chi \mathcal{R}_{\mathrm{U}} \chi '\ \mathrm{iff}\ 
[\forall \mathbf{X} \in \chi (\mathrm{U}),\quad F^0 (\mathbf{X})=x^0\ \mathrm{and}\  \frac{\partial F^j }{\partial x^0}\left(\mathbf{X}\right)=0 \ (j=1,..., N)],
\eea
where $F\equiv \chi '\circ \chi ^{-1}$ is the transition map, whose domain $\chi \left(\mathrm{Dom}(\chi )\cap \mathrm{Dom}(\chi' ) \right) $ contains $\chi (\mathrm{U})$. Then $\mathcal{R}_{\mathrm{U}}$ is an equivalence relation on $\mathcal{A}_\mathrm{U}$. The equivalence classes for this relation are called} reference frames {\it (over the domain $\mathrm{U}$)}.\\

{\it Proof.} i) By definition, $\mathrm{U}\subset \mathrm{Dom}(\chi )$, whence $\chi (\mathrm{U})\subset \chi (\mathrm{Dom}(\chi ))$, for all $\chi \in \mathcal{A}_\mathrm{U}$. It follows that $\forall \chi \in \mathcal{A}_\mathrm{U},\ \chi \mathcal{R}_{\mathrm{U}}\chi $, since $F=\mathrm{Id}_{\chi (\mathrm{Dom}(\chi ))}$ in that case. That is, the relation $\mathcal{R}_{\mathrm{U}}$ is reflexive.\\

ii) If $\chi \mathcal{R}_{\mathrm{U}} \chi '$, let $\mathbf{X'} \in \chi' (\mathrm{U})$, so that the transition map $G\equiv \chi \circ \chi'^{-1} =F^{-1}$ is defined in a neighborhood of $\mathbf{X'}$. We have $\mathbf{X}\equiv G(\mathbf{X'})\in \chi (\mathrm{U})$, and $\mathbf{X'}= F(\mathbf{X})$. In particular, $x'^0=F^0( \mathbf{X})=x^0\equiv G^0(\mathbf{X'})$. Furthermore, we have 
\be
0=\frac{\partial G^j }{\partial x'^\nu }\frac{\partial F^\nu  }{\partial x^0}=\frac{\partial G^j }{\partial x'^0}\frac{\partial F^0 }{\partial x^0}+\frac{\partial G^j }{\partial x'^k}\frac{\partial F^k }{\partial x^0}=\frac{\partial G^j }{\partial x'^0},
\ee
hence $\chi' \mathcal{R}_{\mathrm{U}} \chi $: the relation $\mathcal{R}_{\mathrm{U}}$ is symmetric.\\

iii) If $\chi \mathcal{R}_{\mathrm{U}} \chi '$ and $\chi' \mathcal{R}_{\mathrm{U}} \chi ''$, we have 
\be
\forall {\bf X} \in \chi (\mathrm{U}),\quad  F^0( \mathbf{X})=x^0\
\mathrm{and}\ \frac{\partial F^j }{\partial x^0}\left(\mathbf{X}\right)=0,
\ee
as well as (setting $F'\equiv \chi''\circ \chi'^{-1}$)
\be
\forall {\bf X'} \in \chi' (\mathrm{U}),\quad  F'^0( \mathbf{X'})=x'^0\
\mathrm{and}\ \frac{\partial F'^j }{\partial x'^0}\left(\mathbf{X'}\right)=0,
\ee
and therefore, with $F''\equiv \chi''\circ \chi ^{-1}=F'\circ F$,
\bea\label{x''0=x0}
\forall {\bf X} \in \chi (\mathrm{U}),\quad  {\bf X'}\equiv F({\bf X}) \in \chi' (\mathrm{U})\
\mathrm{and}\ \nonumber\\
F''^0( \mathbf{X})=(F'\circ F)^0( \mathbf{X})=F'^0(F( \mathbf{X}))=F'^0(x^0,(x'^j))=x^0.
\eea
Moreover, we may write 
\bea\label{x''j=x''j(xk)}
\forall {\bf X} \in \chi (\mathrm{U}), \quad \frac{\partial F''^j }{\partial x^0}=\frac{\partial F'^j }{\partial x'^0 }\frac{\partial F^0  }{\partial x^0}+\frac{\partial F'^j }{\partial x'^k }\frac{\partial F^k  }{\partial x^0}=0\times \frac{\partial F^0  }{\partial x^0}+\frac{\partial F'^j }{\partial x'^k }\times 0=0.
\eea
Eqs. (\ref{x''0=x0}) and (\ref{x''j=x''j(xk)}) show that $\chi \mathcal{R}_{\mathrm{U}} \chi ''$: the relation $\mathcal{R}_{\mathrm{U}}$ is also transitive. Q.E.D.\\

\paragraph{Thus, all charts $\chi \in \ $F are defined on an a priori given open set $\mathrm{U}$.}\label{Definition-frame} That is, a reference frame F is an equivalence class of parametrizations for a given open parametrizable subset $\mathrm{U}\subset \mathrm{V}$ of the spacetime $\mathrm{V}$. The equivalence relation between charts is defined by Eq. (\ref{R_U}), thus two charts are equivalent iff i) both are defined on U and ii) inside U, they are related together by a purely spatial change (\ref{purely-spatial-change}). The restriction to charts whose domain contains an a priori given open set U is enough for quantum mechanics in a curved spacetime, in so far as we may suppose that the wave functions $\psi $ have suitable boundary or asymptotic behaviour with respect to the space coordinates of $\mathrm{U}$ in every chart adapted to the reference frame F. Mathematically, one might then assume that the relevant manifold is U and is thus diffeomorphic to ${\sf R}^4$ or to a subset of it. However, no restriction on the spacetime V as a whole has to be imposed.\\

\section{Definition of the associated space manifold M}\label{Definition M}

The three-dimensional space manifold M associated with a reference frame F (the latter term being understood in the sense precised \hyperref[Definition-frame]{just above}) will be, in somewhat imprecise terms,\\

`` the set of the world-lines of the observers bound to F---{\it i.e.,} whose spatial coordinates $x^j$ do not depend on the time $x^0$, in any chart of the class F".\\

\noindent The formal definition of the differentiable manifold M will involve several steps: we will first define the set M, then we will define a topology on M, and finally we will equip this topological space with an atlas of charts. For physics, we will assume $N=3$, but this is by no means necessary in any proof. Since the domain of any chart $\chi \in \mathrm{F}$ contains U, we may and will henceforth restrict those charts to the open set U {\it exactly}. And we put the\\

\paragraph{Definition A.} {\it Let $\mathrm{U}$ be a parametrizable open set in $\mathrm{V}$, and let $\mathrm{F}$ be a class modulo $\mathcal{R}_\mathrm{U}$. A ``world line bound to $\mathrm{F}$" is a subset $l$ of $\mathrm{U}$ such that, for} some $\chi \in \mathrm{F}$,
\be\label{l-in-M}
\exists {\bf x} \in {\sf R}^3:\ l=\chi ^{-1}(\mathrm{I}_{\chi\, {\bf x}}\times \{{\bf x}\}),
\ee
{\it where}
\be\label{I-chi-a}
 \mathrm{I}_{\chi\, {\bf x}}\equiv \{s \in {\sf R}; \ (s,{\bf x}) \in \chi (\mathrm{U}) \}.
\ee

\vspace{3mm}
Note that this definition implies that $l$ is indeed a world line in U, {\it i.e.,} $l$ is the range $l= L(\mathrm{I}_{\chi\, {\bf x}})$ with $\mathrm{I}_{\chi\, {\bf x}}$ an open subset of the real line ${\sf R}$ and $L: s\mapsto \chi ^{-1}(s,{\bf x}) $ a smooth mapping from $\mathrm{I}_{\chi\, {\bf x}}$ to $\mathrm{U}$. The following shows that the statement ``$l$ is bound to F" does not depend on a particular chart $\chi \in\ $F:\\

\paragraph{Proposition A.}\label{PropositionA} {\it Let $\mathrm{F}$ be a class modulo $\mathcal{R}_\mathrm{U}$, and let $l$ be a world line bound to $\mathrm{F}$. Then, for} any $\chi' \in \mathrm{F}$, {\it we have (\ref{l-in-M}), {\it i.e.,} 
\be\label{l-from-chi'}
\exists {\bf x'} \in {\sf R}^3:\ l=\chi'^{-1}(\mathrm{I}_{\chi'\, {\bf x'}}\times \{{\bf x'}\}),
\ee
with [$F\equiv \chi '\circ \chi ^{-1}$ being the transition map satisfying (\ref{R_U})]:
\be\label{x'=f(x)}
{\bf x'}\equiv (F^j({\bf x}))\equiv {\bf f}({\bf x}).\\
\ee

\vspace{3mm}
Proof.} The definition (\ref{l-in-M})--(\ref{I-chi-a}) is equivalent to say that, for some chart $\chi \in \ $F and for some given ${\bf x} \in {\sf R}^3$, we have:
\be\label{l-in-M-simple}
X \in l \Longleftrightarrow \exists s \in {\sf R}: \chi (X)=(s,{\bf x}).
\ee
(In other words, 
\be\label{l-in-M-by-P_S}
X \in l \Longleftrightarrow P_S(\chi (X))={\bf x},
\ee
where $P_S: {\sf R}^4 \rightarrow {\sf R}^3,\ {\bf X}\equiv (x^\mu )\mapsto {\bf x}\equiv (x^j )$, is the spatial projection.) Using the definition (\ref{R_U}) of the equivalence relation between charts, which is satisfied by all pairs $(\chi ,\chi ')$ in the same class F, we may thus write as well, for any other chart $\chi' \in \mathrm{F}$:
\be
X \in l \Longleftrightarrow \exists s \in {\sf R}: \chi' (X)=F(\chi (X))=(s,{\bf x'}), \quad {\bf x'}\equiv (F^j({\bf x})).
\ee
This proves \hyperref[PropositionA]{Proposition A}.\\

We may then define the space M:\\

\paragraph{Definition B.} {\it Let $\mathrm{F}$ be a reference frame, {\it i.e.,} a class modulo $\mathcal{R}_\mathrm{U}$ (\ref{R_U}) for some parametrizable open set $\mathrm{U}\subset \mathrm{V}$. The space $\mathrm{M}$ associated with $\mathrm{F}$ is the set of all world lines bound to $\mathrm{F}$.}\\

\paragraph{Proposition B.}\label{PropositionB} {\it For any event $Y\in \mathrm{U}$, there is a unique world line $l \in \mathrm{M}$ such that $Y \in l$.}\\

{\it Proof.} If there is any such world line $l$, then its definition in a chart $\chi \in\ $F (indeed in any chart $\chi \in\ $F, by \hyperref[PropositionA]{Proposition A}), must be (\ref{l-in-M-simple}) or equivalently (\ref{l-in-M-by-P_S}), with 
\be
{\bf x} \equiv P_S(\chi (Y)).
\ee
Thus, the constant spatial projection ${\bf x}$ is uniquely defined, and so is then the world line (\ref{l-in-M-by-P_S}). Q.E.D.\\

\vspace{3mm}
\paragraph{Definition C.} {\it Let $\mathrm{F}$ be a reference frame, let $l$ be a world line in the associated space $\mathrm{M}$, and let $\chi $ be any chart in the class $\mathrm{F}$. The ``associated chart" is the mapping}
\be\label{def-chi-tilde}
\tilde{\chi }: \mathrm{M}\rightarrow {\sf R}^3,\quad l\mapsto {\bf x}\mathrm{\ with\ }\{{\bf x}\}\equiv P_S(\chi (l)).\\
\ee

\vspace{3mm}
\paragraph{Proposition C.} {\it Consider the set $\mathcal{T}$ of the subsets $\Omega \subset \mathrm{M}$ such that, 
\be\label{def-Topo}
\forall \chi \in \mathrm{F},\quad \tilde{\chi }(\Omega)\mathrm{\ is\ an\ open\ set\ in\ }{\sf R}^3.
\ee
It is a topology on $\mathrm{M}$.}\\

{\it Proof.} i) Clearly, $\emptyset \in \mathcal{T}$. \\

ii) To show that M$\ \in\mathcal{T}$, take any $\chi \in \mathrm{F}$. From the definition (\ref{def-chi-tilde}), it follows that $\tilde{\chi }(\mathrm{M})\subset P_S(\chi(\mathrm{U}))\equiv \mathrm{W}_\chi $. The latter is an open set in ${\sf R}^3$, since {\it a}) U is an open set in V, {\it b}) $\chi $ is bicontinuous, and {\it c}) $P_S$, as a linear mapping of the finite-dimensional vector space ${\sf R}^4$ onto the finite-dimensional vector space ${\sf R}^3$, is an open mapping. Moreover, let ${\bf x}\in\mathrm{W}_\chi $, thus ${\bf x}=P_S(\chi (Y))$ for some $Y \in\ $U. By \hyperref[PropositionB]{Proposition B}, there is a unique world line $l \in \mathrm{M}$ such that $Y \in l$. We have then by (\ref{l-in-M-by-P_S}): $\{{\bf x}\}=P_S(\chi (l))$, that is ${\bf x}=\tilde{\chi }(l)$ by (\ref{def-chi-tilde}). Thus, $\mathrm{W}_\chi \subset\tilde{\chi }(\mathrm{M}) $, whence $\tilde{\chi }(\mathrm{M})= \mathrm{W}_\chi $. Hence M$\ \in\mathcal{T}$ since, for any $\chi \in \mathrm{F}$, $\mathrm{W}_\chi $ is open.\\

iii) Let us show that the intersection of two elements of $\mathcal{T}$ belongs to $\mathcal{T}$. To this aim, we observe that, for any $ \chi \in \mathrm{F}$, $\tilde{\chi }$ is a {\it one-to-one} mapping from M onto $\tilde{\chi }(\mathrm{M})$. Indeed, it follows from (\ref{l-in-M-by-P_S}) that the world line $l \in \mathrm{M}$ is determined uniquely by the data ${\bf x}$, with $\{{\bf x}\}=\tilde{\chi }(l)\equiv P_S(\chi (l))$. Hence, if $\Omega _1$ and $\Omega _2\in\mathcal{T}$, the set $\tilde{\chi }(\Omega _1\cap \Omega _2)=\tilde{\chi }(\Omega _1)\cap \tilde{\chi }(\Omega _2)$ is an open set, for any $\chi \in \ $F, hence $\Omega _1\cap \Omega _2 \in\mathcal{T}$.\\

iv) Finally, from the definition (\ref{def-Topo}), one checks trivially that any union of elements of $\mathcal{T}$ is still an element of $\mathcal{T}$. This completes the proof.\\

\vspace{3mm}
\paragraph{Theorem B.}\label{TheoremB} {\it The set of the ``associated charts": $\tilde{\mathrm{F}}\equiv \{\tilde{\chi };\ \chi \in \mathrm{F}\}$, is an atlas on the topological space $(\mathrm{M},\mathcal{T})$, hence defines a structure of differentiable manifold on $\mathrm{M}$.}\\

{\it Proof.} i) We showed above that, for any $ \chi \in \mathrm{F}$, $\tilde{\chi }$ is a one-to-one mapping from M onto $\tilde{\chi }(\mathrm{M})$. Thus, the inverse mapping $\tilde{\chi }^{-1}$ is well-defined on $\tilde{\chi }(\mathrm{M})$.\\

ii) From the definition (\ref{def-Topo}) of the topology $\mathcal{T}$, it follows immediately that, for any open set W in ${\sf R}^3$, and for any open set W$'\,\equiv W\cap  \tilde{\chi }(\mathrm{M})$ in $\tilde{\chi }(\mathrm{M})$, the inverse range $\Omega \equiv \tilde{\chi }^{-1}(\mathrm{W} )=\tilde{\chi }^{-1}(\mathrm{W}' )$ is an open set in M, $ \Omega \in \mathcal{T}$; and that, conversely, for any open set in M, $ \Omega \in \mathcal{T}$, the inverse range $(\tilde{\chi }^{-1})^{-1}(\Omega )=\tilde{\chi }(\Omega )$ is an open set in $\tilde{\chi }(\mathrm{M})\subset {\sf R}^3$. That is, for any $ \chi \in \mathrm{F}$, the ``associated chart" $\tilde{\chi }$ is a bicontinuous mapping from M onto the open set $\tilde{\chi }(\mathrm{M})\subset {\sf R}^3$, thus is a {\it chart} on M, indeed.\\

iii) Let us show that any two charts $\tilde{\chi },\tilde{\chi' }$ (with $\chi $ and $\chi '\in \ $F) are compatible. That is, let us show that the transition map $\tilde{F}\equiv \tilde{\chi '}\circ \tilde{\chi}^{-1}$, which is defined on $\tilde{\chi}(\mathrm{M})$, is smooth. Let ${\bf x} \in \tilde{\chi}(\mathrm{M})$. Thus, there exists an $l \in \mathrm{M}
$ with $\tilde{\chi}(l)={\bf x}$, {\it i.e.,} we have the equivalence (\ref{l-in-M-by-P_S}). Taking $X_0\in l$, we have thus, for some $s\in {\sf R}$, $\chi (X_0)=(s,{\bf x})\in \chi(\mathrm{M})$. By \hyperref[PropositionA]{Proposition A} and Eq. (\ref{x'=f(x)}), $l$ has also a constant spatial projection ${\bf x'}={\bf f}({\bf x})$ in the chart $\chi '$, so that we get simply from the definition (\ref{def-chi-tilde}):
\be
\tilde{\chi '}\circ \tilde{\chi}^{-1}({\bf x})=\tilde{\chi '}(l)={\bf f}({\bf x})\equiv (F^j({\bf x})).
\ee
Since this is true for any ${\bf x} \in \tilde{\chi}(\mathrm{M})$, it follows that the ``associated", spatial, transition map $\tilde{F}$ is as smooth as is the spacetime transition map $F$.\\

iv) Finally, any single chart $\tilde{\chi}$, associated with some $\chi \in \mathrm{F}$, already covers the whole of the space M, and thus constitutes by itself an atlas of the manifold M. So a fortiori the set $\tilde{\mathrm{F}}\equiv \{\tilde{\chi };\ \chi \in \mathrm{F}\}$, is an atlas. This completes the proof.\\

Note that, from the fact that one chart (in fact, all charts $\tilde{\chi}$) is (are) defined on the whole of the space manifold M, it follows that the topology of M is Hausdorff and 2nd-countable.\\

\section{Discussion}\label{Discussion}

i) As a consequence of Theorems \hyperref[TheoremA]{A} and \hyperref[TheoremB]{B}, we may specify any event $X$ in the open subset U of the spacetime by the corresponding time $t\equiv x^0\equiv P_T(\chi (X))$ and by the spatial position $x\equiv \tilde{\chi }^{-1}(P_S(\chi (X)))\in \mathrm{M}$, thus by the pair $(x^0,x)$. This $1\times 3$ decomposition is {\it independent of the chart} $\chi \in \ $F, since {\it a}) the definition (\ref{R_U}) of the equivalence relation implies that $x^0$ does not depend on the chart $\chi \in \ $F, and {\it b}) by \hyperref[PropositionB]{Proposition B} and definition (\ref{def-chi-tilde}), $x$ is the unique world line $x \in\ $M that contains $X$. In a given chart $\chi \in \ $F, we may also specify the event $X\in \ $U by its coordinate set $(x^\mu )=(x^0,{\bf x})=\chi (X)\in \chi (\mathrm{U})$: the intrinsic and coordinate decompositions are related by ${\bf x}=\tilde{\chi }(x)$. The foregoing results achieve the mathematical justification of the references to the space manifold M in our work about the quantum mechanics of the Dirac equation; see in particular the end of subsect. 4.1. in Ref. \cite{A42}. \\

\vspace{3mm}
\noindent ii) It is striking that the present construction does not need any metrical property of spacetime. Even the choice of the ``time" coordinate $x^0$ is arbitrary: everything would still work if we would give its role to a spatial coordinate $x^j$ instead. In the same way, the Dirac equation makes no formal difference between the time coordinate and the space coordinates. Even the Dirac Hamiltonian, which allows one to rewrite the Dirac equation in the Schr\"odinger form, would have the same general form shown in Ref. \cite{A42} if, after a permutation of the indices, $x^0$  were in fact a spatial coordinate. However, the signature of the spacetime metric appears implicitly in the anticommutation relation satisfied by the Dirac matrices. More specifically, a sufficient condition for the positive-definiteness of the Hilbert space scalar product is that the coordinate system be a physically admissible one in the sense of Eq. (\ref{admissible-coordinates}) \cite{A42}. This condition contains, in particular, the demand that the first coordinate $x^0$ be a time-like coordinate, Eq. (\ref{admissible-coordinates})$_1$. And indeed, for a physically admissible reference frame, the world lines  $l \in \mathrm{M}$, which are just the lines ``$x^0$ variable, $x^j=\,$Constant for $j=1,2,3$", should be world lines of physical observers, thus should be time-like world lines in the spacetime $(\mathrm{V},\,g_{\mu \nu })$. It just turns out that this is not needed in our definition of a reference frame, nor in the construction of the 3-D manifold associated with it. \\

An example is G\"odel's universe, whose line element $\dd s$ is given by \cite{Goedel1949}: 
\be\label{Goedel metric}
\dd s^2 = \left(\dd t + e^{\kappa x} \dd z \right)^2 - \dd x^2-\dd y^2-\frac{1}{2}e^{2\kappa x} \dd z^2,
\ee
where $\kappa$ is a constant parameter.  G\"odel's spacetime manifold is ${\sf R}^4$, so that $(t,x,y,z)$ are just the usual coordinates on ${\sf R}^4$.  Note that both the $t$--coordinate curves and the $z$--coordinate curves are time-like in G\"odel's universe. Therefore, we may interchange the time-like $t$ and $z$ coordinates. After this interchange: $t^\star=z,\ z^\star=t$, which implies moving to another reference frame, the line element (\ref{Goedel metric}) becomes:  
\be\label{Goedel metric- t & z permuted}
\dd s^2 = \left(e^{\kappa x} \dd t^\star +\dd z^\star \right)^2 - \dd x^2-\dd y^2-\frac{1}{2}e^{2\kappa x} \dd t^{\star 2}.
\ee
Let F be the reference frame defined by the chart $\chi: X \mapsto (t,x,y,z)$, whose domain is the whole of ${\sf R}^4$. (In fact, $\chi$ is just the identity map of ${\sf R}^4$.) Thus, F is the set of the charts defined on the whole of ${\sf R}^4$ and which may be exchanged for the chart $\chi$ by a purely spatial change of coordinates (\ref{purely-spatial-change}). Similarly, let F$^\star$ be the reference frame defined by the chart $\chi^\star: X \mapsto (t^\star,x,y,z^\star)$. The 3-D manifolds M and M$^\star$, which we associate with F and F$^\star$ respectively, are both diffeomorphic to ${\sf R}^3$, through the ``associated charts"  $\tilde{\chi}: \mathrm{M} \ni l \mapsto (x,y,z) \in {\sf R}^3$ and $\tilde{\chi}^\star: \mathrm{M}^\star \ni l^\star \mapsto (x,y,z^\star) \in {\sf R}^3$. Using results proved in Ref. \cite{A42}, it can be shown that there is no relevant Hilbert space with a positive scalar product associated with the Dirac equation in G\"odel's universe, neither in the reference frame F nor in the reference frame F$^\star$. This is consistent with the fact that, in view of Eqs. (\ref{Goedel metric}) and (\ref{Goedel metric- t & z permuted}), neither of the coordinates $(t,x,y,z)$ and $(t^\star,x,y,z^\star)$ are admissible in the sense of Eq. (\ref{admissible-coordinates})---since they satisfy the condition (\ref{admissible-coordinates})$_1$ but not the condition (\ref{admissible-coordinates})$_2$. \\

From the components of the spacetime metric $g_{\mu\nu}$, we can define {\it two} metrics on the 3-D manifold M associated with a reference frame:
\bi
\item We can take the induced metric with the negative sign, thus\\
\be\label{spatial-induced}
h_{jk}=-g_{jk}\quad (j,k=1,2,3).   
\ee
\item Or, we can take Landau-Lifshitz and M\o ller's definition---which was shown by them to define the physically relevant spatial distances \cite{L&L,Moeller1952}:
\be\label{spatial-L&L}
h'_{jk}=-g_{jk}+\frac{g_{0j}g_{0k}}{g_{00}}\quad (j,k=1,2,3). 
\ee

\ei 
If we consider the second form (\ref{Goedel metric- t & z permuted}) of the G\"odel metric, corresponding with the new reference frame F$^\star$, the definitions (\ref{spatial-induced}) and (\ref{spatial-L&L}) give us:
\be
(h_{jk})=\begin{pmatrix} 
1 & 0  & 0\\
0  & 1 & 0\\
0 & 0 & -1
\end{pmatrix}; \qquad (h'_{jk})=\begin{pmatrix}
1 & 0  & 0\\
0   & 1 & 0\\
 0 & 0 & 1
\end{pmatrix}.
\ee
We note that the induced metric $h_{jk}$ on the 3-D manifold M$^\star$  associated with the reference frame F$^\star$ is homogeneous, but not  positive-definite. In contrast, the Landau-Lifshitz-M\o ller metric $h'_{jk}$ is homogeneous and isotropic, as well as positive-definite, and in fact $h'_{jk}$ is Euclidean. This example confirms that the Landau-Lifshitz-M\o ller spatial metric (\ref{spatial-L&L}), not the induced metric (\ref{spatial-induced}), is relevant to the spatial distances. 
 \footnote{\
However, the condition that the coordinates are physically admissible, Eq. (\ref{admissible-coordinates}), contains the positive-definiteness of the induced 3-D metric defined in Eq. (\ref{spatial-induced}). It is condition (\ref{admissible-coordinates}) which ensures that 
the Hilbert space scalar product for the Dirac equation is positive-definite \cite{A42}.
}
That is, the inhabitants of G\"odel's universe, if they lived in the reference frame F$^\star$, would discover Euclidean geometry from their measurements of distances and angles, just like Euclid did.
\\

\vspace{3mm}
\noindent iii) In Section \ref{Definition M}, we assumed that the dimension of the spacetime V is $N+1=4$, for an obvious physical reason. However, all proofs, hence all results above, work exactly the same if one substitutes any positive integer $N$ for the particular integer 3, and $N+1$ for 4. Thus, also the dimension of the spacetime can be taken to be an arbitrary integer ($\ge 2$) in the present results.\\

\vspace{3mm}
\noindent iv) Our definition of a reference frame is restricted to a {\it parametrizable} open domain of the spacetime. As we \hyperref[Definition-frame]{outlined}, this seems to be acceptable for the ``practical'' application to quantum mechanics in a curved spacetime. On the other hand, that is not fully satisfactory from the viewpoint of the global topology of spacetime---although we did not need to assume anything about it for the present results. However, it does not seem possible to define a relevant equivalence relation between charts, as the relation (\ref{R_U}), if the different charts are not defined in a common domain U. 

%\newpage
%\bigskip
%\newpage

\end{document}